\documentclass[twocolumn]{emulateapj}

\usepackage{graphicx}
\usepackage{epsfig}

\begin{document}

\title{MASSES OF NEARBY SUPERMASSIVE BLACK HOLES WITH VERY-LONG BASELINE INTERFEROMETRY}
\shorttitle{MASSES OF NEARBY SUPERMASSIVE BLACK HOLES WITH VLBI}
\shortauthors{JOHANNSEN ET AL.}

\author{Tim Johannsen\altaffilmark{1}, Dimitrios Psaltis\altaffilmark{1}, Stefan Gillessen\altaffilmark{2}, Daniel P. Marrone\altaffilmark{1}, Feryal \"Ozel\altaffilmark{1}, \\
Sheperd S. Doeleman\altaffilmark{3}, and Vincent L. Fish\altaffilmark{3}}
\affil{$^1$ Physics and Astronomy Departments, University of Arizona, 933 N. Cherry Avenue, Tucson, AZ 85721, USA\\
$^2$ Max Planck Institut f\"ur Extraterrestrische Physik, Giessenbachstra\ss e, D-85748 Garching, Germany\\
$^3$ Massachusetts Institute of Technology, Haystack Observatory, Route 40, Westford, MA 01886, USA}
\email{(TJ) timj@physics.arizona.edu}

\begin{abstract}

Dynamical mass measurements to date have allowed determinations of the mass $M$ and the distance $D$ of a number of nearby supermassive black holes. In the case of Sgr~A*, these measurements are limited by a strong correlation between the mass and distance scaling roughly as $M \sim D^2$. Future very-long baseline interferometric (VLBI) observations will image a bright and narrow ring surrounding the shadow of a supermassive black hole, if its accretion flow is optically thin. In this paper, we explore the prospects of reducing the correlation between mass and distance with the combination of dynamical measurements and VLBI imaging of the ring of Sgr~A*. We estimate the signal to noise ratio of near-future VLBI arrays that consist of five to six stations, and we simulate measurements of the mass and distance of Sgr~A* using the expected size of the ring image and existing stellar ephemerides. We demonstrate that, in this best-case scenario, VLBI observations at 1~mm can improve the error on the mass by a factor of about two compared to the results from the monitoring of stellar orbits alone. We identify the additional sources of uncertainty that such imaging observations have to take into account. In addition, we calculate the angular diameters of the bright rings of other nearby supermassive black holes and identify the optimal targets besides Sgr~A* that could be imaged by a ground-based VLBI array or future space-VLBI missions allowing for refined mass measurements.

\end{abstract}

\keywords{accretion, accretion disks --- black hole physics --- galaxy: center --- galaxies: nuclei --- gravitational lensing: strong}

\section{INTRODUCTION}

Sgr~A*, the supermassive black hole at the center of our galaxy, has been observed for several decades. Monitoring stars orbiting around Sgr~A* has led to measurements of its mass and distance (Ghez et al. 2008; Gillessen et al. 2009). However, these measurements of mass and distance are strongly correlated. For purely astrometric measurements, mass and distance are related as $M \sim D^3$, while for measurements of radial velocities mass and distance are related as $M \sim D^0$. For combined data sets, the correlation between mass and distance behaves roughly as $M \sim D^2$ (Ghez et al. 2008; Gillessen et al. 2009). This correlation between mass and distance constitutes a major source of uncertainty in our understanding of the properties of Sgr~A*. Likewise, dynamical measurements of the masses of a number of nearby supermassive black holes have been obtained with often much greater uncertainties (see, e.g., G\"ultekin et al. 2009).

Another technique, VLBI, aims to image Sgr~A* directly. Recent VLBI observations with an array consisting of the Submillimeter Telescope Observatory (SMTO) in Arizona, the James Clerk Maxwell Telescope (JCMT) on Mauna Kea, and several of the dishes of the Combined Array for Research in Millimeter-wave Astronomy (CARMA) in California resolved Sgr~A* on scales comparable to its event horizon and identified sub-horizon size structures (Doeleman et al. 2008; Fish et al. 2011). Images of accretion flows around black holes have the shadow of the compact object imprinted on them, which depends uniquely on its mass, spin, and inclination (e.g., Falcke et al. 2000) as well as on possible deviations from the Kerr metric (Johannsen \& Psaltis 2010). Based on such images and assuming the mass and distance obtained from the monitoring of stellar orbits, these VLBI observations inferred constraints on the inclination and spin of Sgr~A* (Broderick et al. 2009, 2011) and placed limits on potential non-Kerr signatures (Broderick et al. 2012).

In addition to the shadow, images of optically thin accretion flows around black holes carry a characteristic signature in the form of a bright ring (Johannsen \& Psaltis 2010), which we refer to as the photon ring. Light rays that approach the event horizon closely orbit around the black hole many times before they are detected by a distant observer, resulting in a bright ring due to their long optical path length through the accretion flow. The flux of such photons can account for a significant fraction of the total disk flux and produce higher order images (Cunningham 1976; Laor, Netzer, \& Piran 1990; Viergutz 1993; Bao, Hadrava, \& {\O}stgaard 1994; ${\rm \check{C}ade\check{z}}$, Fanton, \& Calvani 1998; Agol \& Krolik 2000; Beckwith \& Done 2005). These photon rings are clearly visible in all time-dependent general-relativistic simulations of accretion flows that have been reported to date (Mo$\acute{\rm s}$cibrodzka et al. 2009; Dexter, Agol, \& Fragile 2009; Shcherbakov \& Penna 2010).

Johannsen \& Psaltis (2010) showed that a measurement of the ring diameter measures the ratio $M/D$ for the black hole, independent of its spin or deviation from the Kerr metric. Therefore, combining such a measurement with the observations of stars around Sgr~A* can reduce the correlation between mass and distance.

In this paper, we explore the ability of this approach to refine the mass and distance measurements of Sgr~A*. We estimate the precision with which a thermal noise-limited VLBI array can infer the diameter of the ring of Sgr~A* and use a Bayesian technique to simulate measurements of the mass and diameter of Sgr~A* in conjunction with parameters inferred from the existing data of the orbits of stars at comparable wavelengths. We show that, in this best-case scenario, the correlation between mass and distance is reduced significantly. In addition, we argue that the accretion flows of other nearby supermassive black holes are optically thin, allowing for VLBI observations of their respective photon rings. We assess the prospects of using this technique to infer the masses of these sources.

\section{MEASURING THE PHOTON RING OF SGR A*}

The properties of photon rings are practically independent of the specific flow geometry and remain constant even if the accretion flow itself is highly variable (Johannsen \& Psaltis 2010). The relative brightness as well as the constancy of these rings make them ideal targets for VLBI-imaging observations.

For a Kerr black hole with mass $M$, the shape of a given photon ring has a diameter of
\begin{equation}
L \simeq 10.4~r_g,
\label{ringdiameter}
\end{equation}
which remains practically constant for all values of the spin and disk inclination (Johannsen \& Psaltis 2010). In this expression, 
\begin{equation}
r_g \equiv \frac{GM}{c^2}
\end{equation}
is the gravitational radius, and $G$ and $c$ are the gravitational constant and the speed of light, respectively.

The angular diameter $\theta_{\rm ring}$ of the diameter of the photon ring of a black hole is given by the ratio of its diameter and distance,
\begin{equation}
\theta_{\rm ring}=\frac{L}{D}.
\label{openingangle}
\end{equation}
Assuming the current mass and distance measurements of Sgr~A*, $M_0=4.3\times10^6M_{\odot}$ and $D_0=8.3~{\rm kpc}$ (Gillessen et al. 2009), the photon ring has an angular diameter of
\begin{equation}
\theta_0 \simeq 53~\mu{\rm arcsec}.
\end{equation}

\begin{figure}[ht]
\begin{center}
\psfig{figure=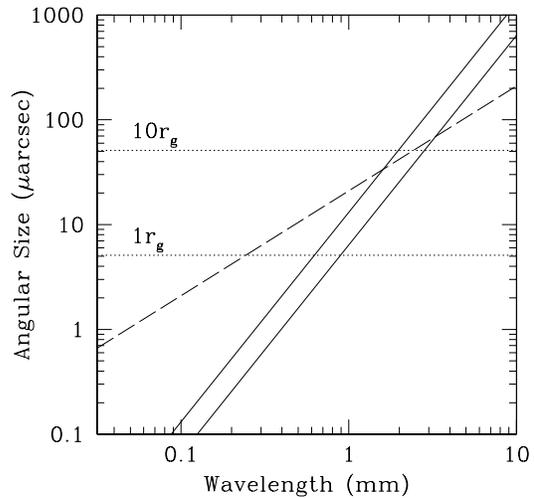,height=3in}
\end{center}
\caption{{\em Solid lines:} Degree of the blurring of structures on the image of Sgr~A*, inferred using the interstellar scattering law of Bower et al. (2006); the two lines correspond to the major and minor axes of the scattering ellipse. {\em Long dashed line:} Resolution estimate of a radio interferometer, taken as $\theta_{\rm res}=k\lambda/d$ with $k=1$ and $d=10^4{\rm km}$. {\em Dotted lines:} Angular diameter of Sgr~A* corresponding to length scales of $10r_g$ and $1r_g$, respectively. Imaging the photon ring of Sgr~A* with VLBI is possible at wavelengths $\lambda\lesssim1{\rm mm}$.}
\label{f:openingangle}
\end{figure}

Radio interferometers are limited by their intrinsic resolution as well as by interstellar scattering. In order to identify the range of wavelengths within which VLBI measurements of the photon ring of Sgr~A* are resolution-limited, we compare the blurring effects of interstellar scattering with the resolution of an interferometer. In Figure~\ref{f:openingangle} we plot the minimum size of resolvable structures on the image of Sgr~A* using the interstellar scattering law of Bower et al. (2006). We also estimate (dashed line) the resolution of a radio interferometer at a given wavelength $\lambda$ by the expression
\begin{equation}
\theta_{\rm res} = k\lambda/d
\label{resolution}
\end{equation}
 with $k=1$ and a diameter $d=10^4{\rm km} \sim d_{\rm earth}$, which is comparable to the baseline length between the JCMT on Hawaii and the South Pole Telescope. This yields $\theta_{\rm res, \mu as}=21\lambda_{\rm mm}$, where $\theta_{\rm res, \mu as}$ is the resolution in $\mu$arcsec and $\lambda_{\rm mm}$ is the observed wavelength in mm. Dotted lines mark the angular diameters corresponding to length scales of $10r_g$ and $1r_g$ at the assumed distance of Sgr~A*. As can be seen from this figure, at sub-mm wavelengths interstellar scattering becomes negligible and measurements are limited by the resolution. Therefore, a measurement of the diameter of the photon ring of Sgr~A* will require VLBI observations at wavelengths $\lambda\lesssim1{\rm mm}$.

In the following, we estimate the observed flux of the photon ring of Sgr A* assuming a Schwarzschild black hole and employing a model of a geometrically thin advection-dominated accretion flow (ADAF; Narayan \& Yi 1994, 1995; Narayan, Yi, \& Mahadevan 1995). ADAFs model the accreting gas around a supermassive black hole as a quasi-spherically symmetric plasma consisting of thermal electrons and ions at different temperatures (Narayan \& Yi 1995a,~b) as well as of non-thermal electrons (Mahadevan 1999; \"Ozel, Psaltis, \& Narayan 2000; Yuan, Quataert, \& Narayan 2003). Such an accretion flow is allowed to cool through comptonization and the emission of bremsstrahlung and synchrotron radiation, with the latter generating the predominant contribution to the observed spectrum of Sgr A* at radio and \mbox{sub-mm} wavelengths (Narayan et al. 1995).

For our estimate, we follow Broderick et al. (2009) and assume an ADAF model with a density of thermal electrons
\begin{equation}
n_e(r)=n_{e0}\left(\frac{r}{r_g}\right)^{-1.1},
\label{edensity}
\end{equation}
electron temperature
\begin{equation}
T_e(r)=T_{e0}\left(\frac{r}{r_g}\right)^{-0.84},
\label{temperature}
\end{equation}
and magnetic field
\begin{equation}
\frac{B^2}{8\pi}=\beta^{-1}n_e \frac{2G M_0 m_p}{12r},
\label{Bfield}
\end{equation}
where $\beta = 10$. For the coefficients we use (Broderick et al. 2011)
\begin{equation}
n_{e0}=3\times10^7~{\rm cm}^{-3}
\end{equation}
and
\begin{equation}
T_{e0}=1.7\times10^{11}~{\rm K}.
\end{equation}
These coefficients lead to predictions of the spectrum, polarization, and image size for Sgr~A* that are in agreement with all current observations.

We assume that all the emission at mm-wavelengths is due to thermal synchrotron radiation (see Narayan et al. 1995). While there may be small contributions of synchrotron emission from non-thermal electrons (e.g., Mahadevan 1998; \"Ozel et al. 2000; Yuan et al. 2003) or of other types of radiation at these wavelengths (such as jets; e.g., Falcke et al. 1993), this assumption affects our analysis only marginally. In practice, the measured total flux incorporates all such contributions.

Following Dolence et al. (2009), we write the synchrotron emissivity as
\begin{equation}
j_\nu\simeq\frac{ \sqrt{2}\pi e^2 n_e \nu_s }{ 3c K_2(1/\theta_e) }(X^{1/2}+2^{11/12}X^{1/6})^2 \exp(-X^{1/3}),
\label{dolence}
\end{equation}
where
\begin{equation}
X\equiv\frac{\nu}{\nu_s},
\end{equation}
\begin{equation}
\nu_s\equiv\frac{2}{9}\left(\frac{eB}{2\pi m_e c}\right)\theta_e^2\sin\alpha,
\end{equation}
\begin{equation}
\theta_e\equiv\frac{kT}{m_e c^2},
\end{equation}
and $K_2$ is the modified Bessel function of the second kind of the second order. Here, $\alpha$ is the angle between the wave vector of the emitted photon and the magnetic field. For our estimate, we use the average $<\sin\alpha>= \pi/4$.

We assume that optical paths follow at least one loop along the circular photon orbit of a Schwarzschild black hole located at radius $r=3r_g$. The emitted intensity of radiation is then given by the expression
\begin{equation}
I_{\rm em}=6\pi r_g j_\nu.
\end{equation}
The observed intensity is related to the emitted intensity by the third power of the redshift factor, which we take to be the gravitational redshift of a photon observed at infinity emitted from $r=3r_g$ (i.e., we neglect for this simple estimate the high velocity of the flow, which will serve to increase the intensity at infinity). This yields:
\begin{equation}
I_{\rm obs}=(1+z)^{-3}I_{\rm em},
\end{equation}
where
\begin{equation}
1+z=\frac{1}{\sqrt{1-\frac{2}{3}}}=\sqrt{3}.
\end{equation}

The photon ring has an approximate width of
\begin{equation}
\Delta L \simeq 0.1r_g \cong 0.51~{\rm \mu arcsec}
\end{equation}
in the image plane (Johannsen \& Psaltis 2010). Then, the subtended solid angle is given by the expression
\begin{equation}
\Omega=\frac{L\times \Delta L}{D_0^2} \simeq \frac{\pi r_g^2}{D_0^2}.
\end{equation}

This yields our estimate for the observed flux density of the photon ring:
\begin{equation}
F_{\nu,{\rm ring}}=I_{\rm obs}\times\Omega.
\end{equation}

\begin{figure}[t]
\begin{center}
\psfig{figure=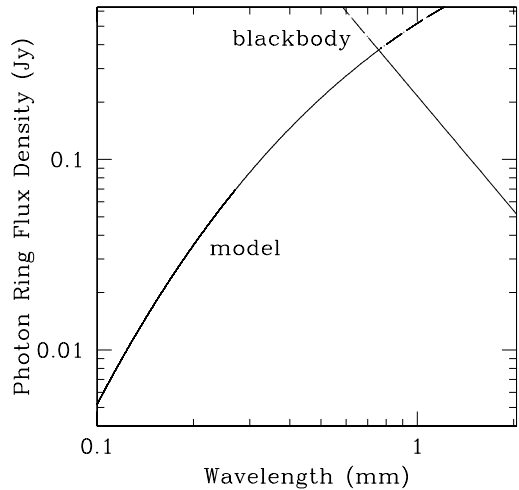,height=3in}
\end{center}
\caption{Estimate of the flux density of the photon ring of Sgr~A$^*$ assuming emission from a geometrically thin ADAF around a Schwarzschild black hole (curve labeled ``model''). At wavelengths $\lambda\lesssim 0.8~{\rm mm}$, the modeled flux density is of the order of 0.2~Jy. We also plot (curve labeled ``blackbody'') a blackbody function at the same temperature and emission radius, which sets an upper flux density limit to the ring emission of the ADAF model at wavelengths $\lambda\gtrsim 0.8~{\rm mm}$. Beyond this wavelength, the emission becomes optically thick. The solid line marks our estimate of the ring flux density in both wavelength ranges corresponding to the minimum of both functions.}
\label{ringflux}
\end{figure}

In Figure~\ref{ringflux}, we plot the modeled ring flux density as a function of wavelength. We also plot a blackbody function evaluated at the same emission radius and temperature given by expression (\ref{temperature}) as an upper flux density limit. The above estimate exceeds the blackbody flux density at wavelengths $\lambda\gtrsim 0.8~{\rm mm}$, and, thus, at these longer wavelengths, the emission becomes self-absorbed. Therefore, we use the minimum of these two flux densities as an estimate of the ring flux density. At wavelengths $\lambda\lesssim 0.8~{\rm mm}$, the modeled flux density of the photon ring of Sgr~A* is $\sim0.2~{\rm Jy}$, about 1/15 of the total source flux density. Since $\lambda\lesssim 0.8~{\rm mm}$ is also in the regime where VLBI observations are resolution-limited (see Figure~\ref{f:openingangle}), this range of wavelengths is optimal for such measurements.

\begin{figure*}[t]
\begin{center}
\psfig{figure=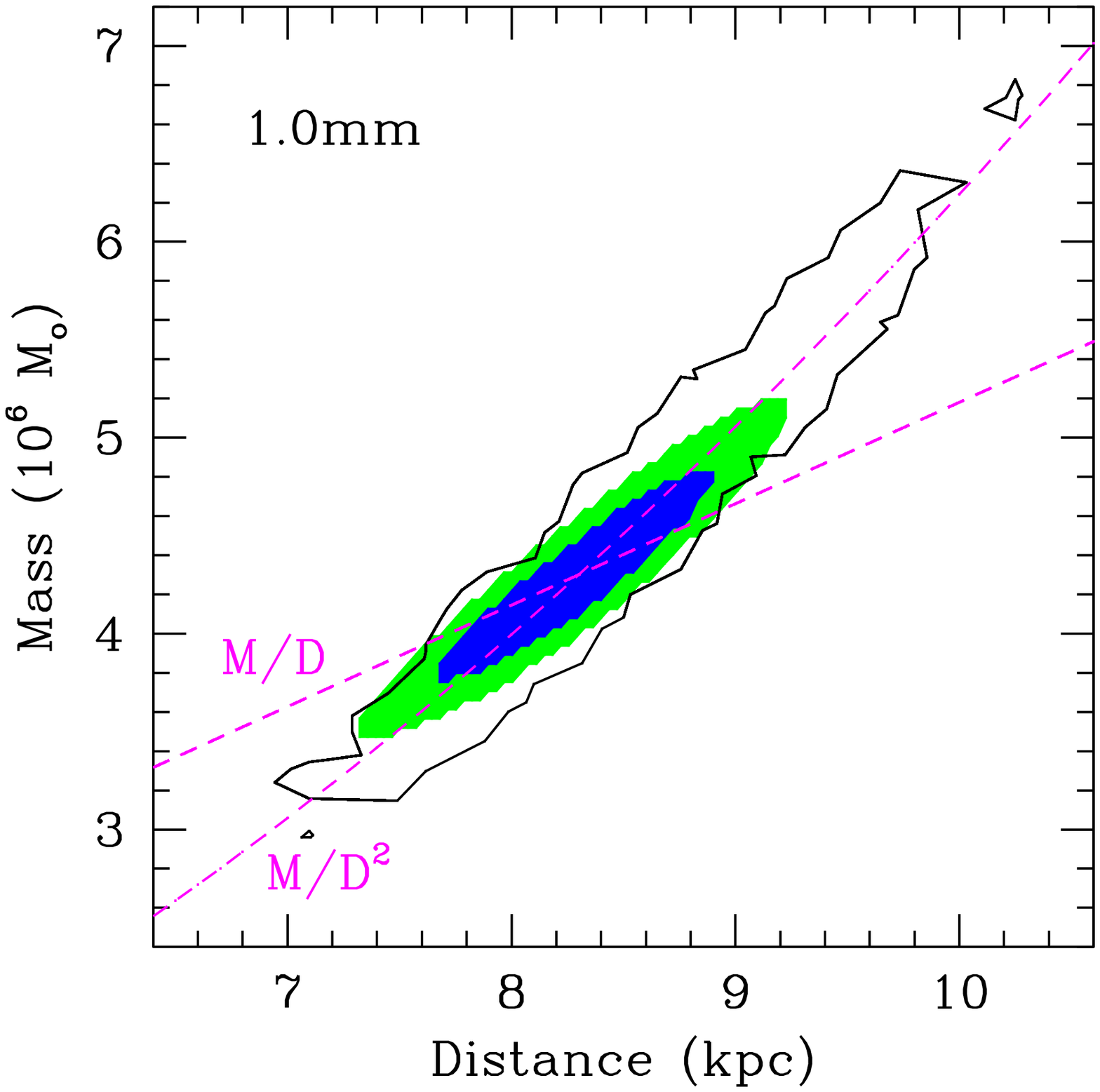,height=2.3in}
\psfig{figure=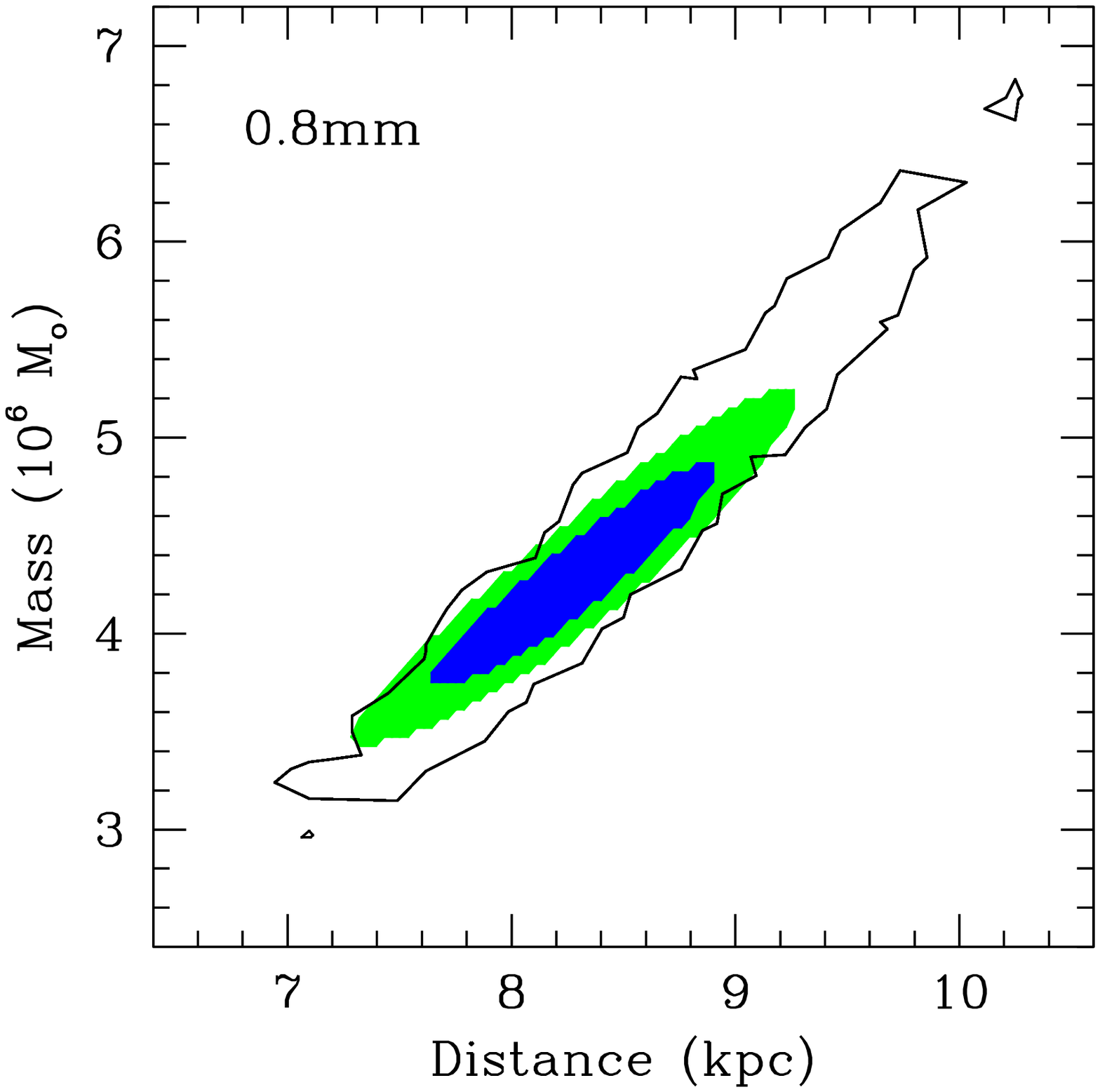,height=2.3in}
\psfig{figure=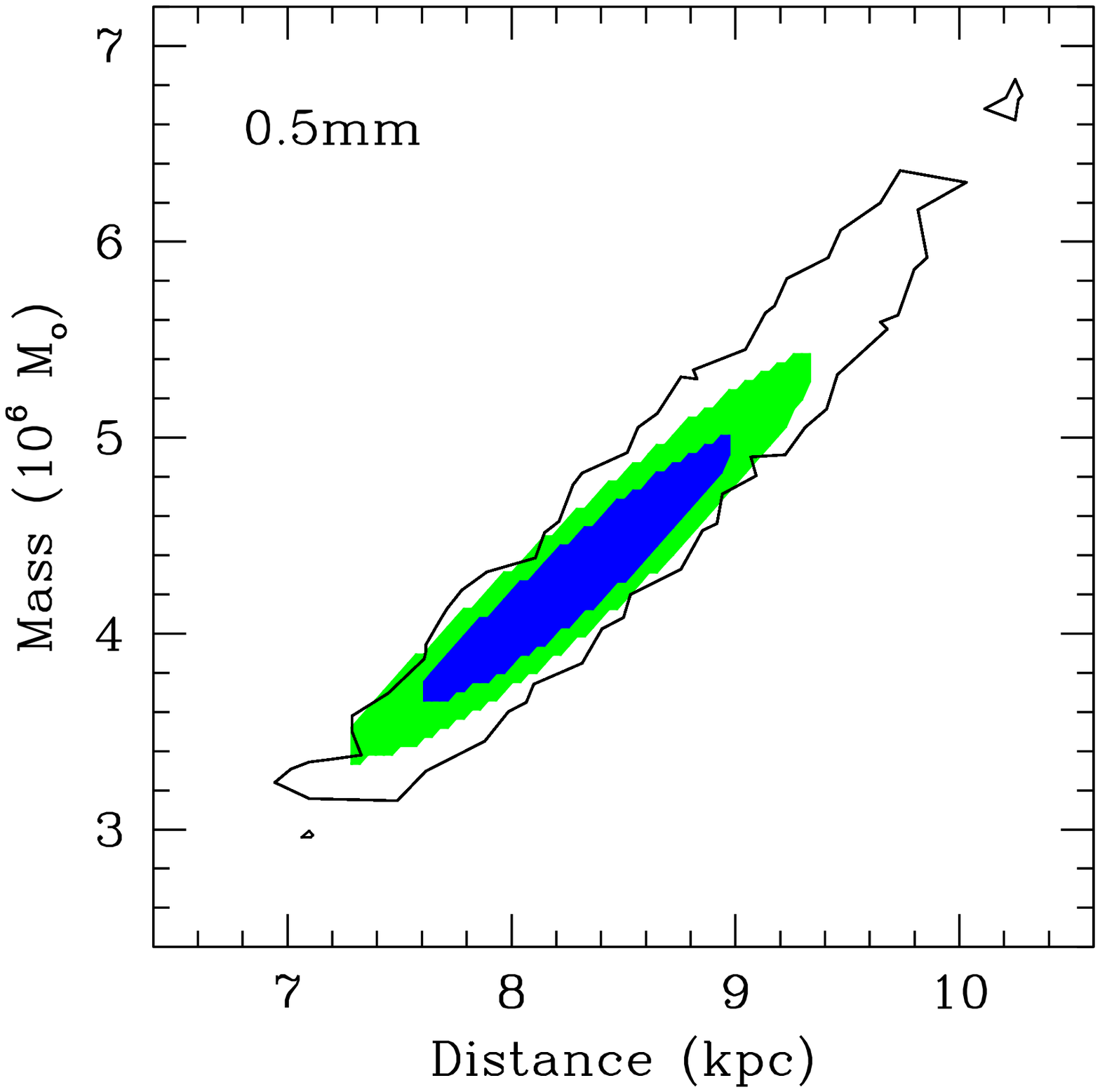,height=2.3in}
\end{center}
\caption{68\% and 95\% confidence contours of the mass and distance of Sgr~A* for the combined distribution of stellar orbits and simulated VLBI measurements for a thermal noise-limited array at several wavelengths. Compared to (solid line) the 95\% confidence contour of dynamical observations alone (Gillessen et al. 2009), the improvement of the mass and distance measurements is similar at all three wavelengths. For comparison, we also plot (left panel, dashed lines) the constant ratios $M/D$ and $M/D^2$ in order to illustrate the dependence of both individual methods on the correleation between mass and distance.}
\label{contours}
\end{figure*}

The ring diameter is determined solely by the mass via equation (\ref{ringdiameter}), and we can relate the black-hole mass and the angular diameter of the ring according to the expression
\begin{equation}
\frac{M_0}{10^6M_{\odot}}=9.8\times10^{-3}\frac{\theta_{\rm ring}}{\rm \mu arcsec}\frac{D_0}{\rm kpc}.
\end{equation}
Therefore, for VLBI imaging, the mass is proportional to the distance, and we can reduce the correlation between mass and distance in combination with dynamical measurements.

We now explore the prospect of combining dynamical measurements of Sgr~A* with VLBI imaging observations of the photon ring. We analyze the best-case scenario of a thermal noise-limited VLBI array in order to assess whether such a measurement is worthwhile. Systematic limitations will degrade the VLBI measurement somewhat, as discussed below.

We employ a Bayesian analysis to estimate the probability distribution over the mass and distance of Sgr~A* from a measurement of the angular diameter of the photon ring in combination with the constraints obtained from stellar dynamics. We take the latter as our prior, $P_{\rm prior}(M,D)$, by converting into a likelihood the $\chi^2$ distribution with $\nu=114$ degrees of freedom obtained from the existing data set of the ephemerides of several S-stars (Gillessen et al. 2009). We assume a Gaussian posterior likelihood of obtaining a particular measurement of the angular diameter as
\begin{equation}
P_{\rm VLBI}({\rm data}|M,D) = \frac{1}{\sqrt{ 2\pi \sigma^2 }}\exp \left[ { -\frac{ (\theta_{\rm ring} - \theta_0)^2 }{ 2\sigma^2 } } \right],
\end{equation}
where 
$\theta_{\rm ring}$ is given by equation (\ref{openingangle}) and $\theta_0 = 53~\mu{\rm arcsec}$. We then use Bayes' theorem to write the likelihood of a particular mass and distance of the black hole given the data as
\begin{equation}
P(M,D|{\rm data}) = C P_{\rm VLBI}({\rm data}|M,D) P_{\rm prior}(M,D),
\end{equation}
where $C$ is the appropriate constant that normalizes the likelihood.

The measurement uncertainty $\sigma$ is the key parameter of the likelihood $P_{\rm VLBI}$. In the following, we estimate the scaling of the measurement uncertainty with resolution. The spatial frequencies of interest are those beyond the first null of the Bessel function describing the Fourier transform of the ring structure in visibility space. For a given resolution and ring size, there are $N$ accessible half periods of the oscillation, where $N$ is given by
\begin{equation}
N \equiv \frac{\theta_0}{\theta_{\rm res}} - 1 = \frac{53}{21\lambda_{\rm mm}} - 1.
\end{equation}
The typical amplitude of a Bessel function oscillation of the first few maxima is of order 0.3 of the peak. Therefore, the signal we seek to measure has an amplitude of
\begin{equation}
A \equiv 0.3 \times \frac{F_{\nu,{\rm ring}}}{F_{\nu,{\rm flow}}} \times N = 0.3 \times \frac{0.2}{3} \times N = 0.02 N,
\end{equation}
where $F_{\nu,{\rm flow}} \simeq 3~{\rm Jy}$ is the observed flux density of Sgr~A* near 1~mm (see, e.g., Broderick et al. 2009).

We estimate the expected signal to noise ratio on this measurement by scaling it from the current VLBI measurements. Fish et al. (2011) measure the size of Sgr~A* with a signal to noise ratio of $\sim40$. Near-future VLBI arrays will incorporate 5 or 6 stations. One of them, ALMA, will have the sensitivity of 50 of the current stations resulting in an overall increase in array sensitivity of a factor of up to roughly 9.3. In addition, the scheduled increase of recording bandwidth will increase the sensitivity by a factor of $\sqrt{8}$. To account for the variation in system temperature for typical observing conditions and receiver performance we introduce a degradation in SNR proportional to the observing wavelength squared, normalized to the 1.3~mm performance in Fish et al. (2011). The total signal to noise ratio of this measurement will then be
\begin{eqnarray}
{\rm SNR} &=&  9.3 \times \sqrt{8} \times 40 \times A \times \left( \frac{ \lambda }{ {\rm 1.3~mm} } \right)^2 \nonumber \\
&& \simeq 12 \times \lambda^2_{\rm mm} \left( \frac{53}{21\lambda_{\rm mm}} - 1 \right),
\label{SNR}
\end{eqnarray}
with the width of the distribution given by the expression
\begin{equation}
\sigma = \frac{\theta_0}{{\rm SNR}} \simeq 4.3 \times \lambda^{-2}_{\rm mm} \left( \frac{53}{21\lambda_{\rm mm}} - 1 \right)^{-1}~{\rm \mu arcsec}.
\label{distwidth}
\end{equation}

In Figure~\ref{contours}, we plot confidence contours of the joint probability distribution for the combination of future thermal noise-limited VLBI and current astrometric observations at three different wavelengths. The solid line marks the 95\% confidence contour determined by the stellar ephemerides (Gillessen et al. 2009). A VLBI measurement at a wavelength of $1~{\rm mm}$ significantly improves the result from stellar orbits alone. At smaller wavelengths, the constraints on the mass and distance of Sgr~A* are similar. In the rightmost panel of Figure~3, we have extrapolated the distribution width $\sigma$ given by expression (\ref{distwidth}) to a nominal wavelength of $\lambda=0.5~{\rm mm}$. Measurements at such short wavelengths will be limited by weather conditions and may have to rely on a smaller array with fewer telescopes.

Real observations will face more stringent limitations than those imposed by the interferometer thermal noise due to the complications of astrophysics and measurement systematics. The chief astrophysical limitation is the separation of the ring emission from the source structure in the uv-plane. In our estimate, we have used the location of the nulls as a benchmark for the uncertainties we expect from the VLBI measurement. In practice, however, the full visibility function has to be analyzed with a pattern matching technique that identifies the structure of the ring. Such a technique has to extract the ring from a uv-plane that is only partially sampled by a given set of baselines.

The physics of the accretion flow will also complicate things, as the structure of Sgr A* may vary over the course of an observation. However, because the ring structure is persistent and only weakly altered by rapid changes in the accretion flow we expect that temporal averaging of the visibilities across multiple observing epochs will diminish the importance of such changes.

The VLBI measurement itself must surmount systematic limitations to make the moderate dynamic range measurements proposed here. Chief among these is the difficulty of calibrating the noise level at individual stations, which imposed a 5\% uncertainty in Fish et al. (2011) with the three-station array. In observations with the larger array considered here, there will be many more internal cross-checks available to improve the relative calibration of stations (the absolute calibration is not important). In particular, the use of three phased interferometers (Mauna Kea, CARMA, ALMA) that simultaneously record conventional interferometric data will permit scan-by-scan cross calibration of the amplitude scale of the array.

Furthermore, the larger arrays will be able to make use of closure phases and closure amplitudes that are immune to calibration errors as part of the ring detection, although we have ignored such procedures here because of the difficulty of simply parameterizing the improvement they can permit. Other effects, such as the coherence of the reference systems between stations (reported as $<5$\% in Fish et al. 2011) can be more carefully measured and corrected to prevent them from imposing fundamental limitations to the ring detection.

\section{Other Sources}

Besides Sgr~A*, there exist other nearby supermassive black holes, for which a combination of dynamical measurements and VLBI observations could be feasible. Since these supermassive black holes are located in host galaxies other than the Milky Way, observations are much less affected by interstellar scattering. As an example, Broderick \& Loeb (2009) and Takahashi \& Mineshige (2011) analyzed the prospects of imaging the shadow of M87 with VLBI observations at several different wavelengths.

\begin{figure}[h]
\begin{center}
\psfig{figure=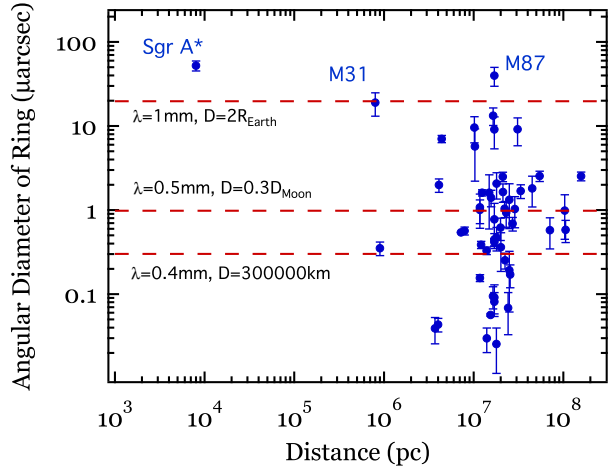,height=2.5in}
\end{center}
\caption{Angular diameters and distances of several supermassive black holes. Sgr~A* has the largest angular diameter, closely followed by M87 due to its high mass, making these sources ideal targets for VLBI imaging. Data taken from G\"ultekin et al. (2009).}
\label{SMBHs}
\end{figure}

In Figure~\ref{SMBHs}, we plot the angular diameter of the photon rings against the distances of a collection of nearby supermassive black holes. Sgr~A* is closest to us and has the largest angular diameter, closely followed by M87 and M31 due to their large black hole masses. The top dashed line indicates the resolution of a telescope array with a baseline equal to the diameter of the Earth (from equation (\ref{resolution})) at a wavelength of 1~mm. For comparison, we also show the resolution of a future space telescope located at 30\% the distance to the moon (comparable to the orbit of {\em Chandra}) at a wavelength of 0.5~mm corresponding to an angular diameter of about $1~{\rm \mu arcsec}$ as well as of the proposed {\em Millimetron} mission (Wild et al. 2009) at a distance of $3\times10^5~{\rm km}$ and a wavelength of $\lambda = 0.4~{\rm mm}$.

\begin{deluxetable*}{lcccccccc}
\tabletypesize{\scriptsize}
\tablecolumns{9}
\tablewidth{0pt}
\tablecaption{Sources for VLBI Observations}
\tablehead{
	\colhead{Source}&
	\colhead{$\theta_{\rm ring}$}&
	\colhead{Distance\tablenotemark{a}}&
	\colhead{$\log(M_{BH})$\tablenotemark{a}}&
	\colhead{$\log(L_R)$\tablenotemark{a}}&
	\colhead{$S_\nu$}&
	\colhead{Ref}&
	\colhead{$\log(L_{\rm bol}/L_{\rm Edd})$\tablenotemark{b}}&
	\colhead{$\lambda_{\rm max}$}\\
   & ($\mu{\rm arcsec}$) & (Mpc) & $(M_\odot)$ & (erg/s) & (Jy) &  &  & (mm) }

	\startdata
Sgr A* & 53 & 0.008 & $6.61\pm0.064$ & 32.48 & 2.4 & 1 & ... & $\sim 1$\tablenotemark{c} \\
NGC 4486 (M87) & 22 & 17.0 & $9.56\pm0.126$ & 39.83 & 0.897 & 2 & -5.15 & $\sim 1$\tablenotemark{d} \\
NGC 0224 (M31) & 19 & 0.8 & $8.17\pm0.161$ & 32.14 & $\approx 3 \times 10^{-5}$ & 3 & -8.90 &  \\
NGC 4649 (M60) & 13 & 16.5 & $9.33\pm0.117$ & 37.45 & $<0.004$ & 4 & -7.84 & $\sim 3$\tablenotemark{d} \\
NGC 3115 & 9.6 & 10.2 & $8.98\pm0.182$ & ... & ... & & -7.03 &  \\
IC 1459 & 9.2 & 30.9 & $9.44\pm0.196$ & 39.76 & 0.264 & 5 & ... &  \\
NGC 4374 (M84) & 9.1 & 17.0 & $9.18\pm0.231$ & 38.77 & 0.129 & 6 & -6.29 &  \\
NGC 5128 (Cen A) & 7.0 & 4.4& $8.48\pm0.044$ & 39.85 & 6.9 & 7 & ... &  \\
NGC 4594 (M104) & 5.7 & 10.3 & $8.76\pm0.413$ & 37.89 & 0.25 & 8 & -4.68 & \\
IC 4296 & 2.5 & 54.4 & $9.13\pm0.065$ & 38.59 & 0.155 & 9 & ... &  \\
NGC 1399 & 2.5 & 21.1 & $8.71\pm0.060$ & ... & $\approx 0.04$ & 3 & ... & $\sim 3$\tablenotemark{d} \\
NGC 4342 & 2.1 & 18.0 & $8.56\pm0.185$ & ... & ... & & ... & \\
NGC 3031 (M81) & 2.0 & 4.1 & $7.90\pm0.087$ & 36.97 & 0.1812 & 10 & -5.29 & $\sim 10$\tablenotemark{e} \\
NGC 4261 & 1.7 & 33.4 & $8.74\pm0.090$ & 39.32 & 0.059 & 11 & -5.21 & \\
NGC 3585 & 1.6 & 21.2 & $8.53\pm0.122$ & ... & ... & & ... &  \\
NGC 3998 & 1.6 & 14.9 & $8.37\pm0.431$ & 38.03 & $<0.007$ & 8 & -4.43 \\
NGC 4697 & 1.6 & 12.4 & $8.29\pm0.038$ & ... & $<0.007$ & 8 & ... & \\
NGC 4026 & 1.4 & 15.6 & $8.33\pm0.109$ & ... & ... & & ... &  \\
NGC 3379 (M105) & 1.1 & 11.7 & $8.09\pm0.250$ & 35.81 & ... & & -7.57 & \\
NGC 3245 & 1.0 & 22.1 & $8.35\pm0.106$ & 36.98 & ... & & -5.83 & \\
NGC 5845 & 1.0 & 28.7 & $8.46\pm0.223$ & ... & ... & & ... & \\
NGC 3377 & 1.0 & 11.7 & $8.06\pm0.163$ & ... & ... & & -6.16 & \\
	\enddata
\tablerefs{ (a)~G\"ultekin et al. (2009); (b)~L.~Ho, priv. commun.; (c)~Broderick et al. (2009); (d)~Di Matteo et al. (2000); (e)~Doi et al. (2005); For $S_\nu$: (1)~VLBI:~1.3~mm (Doeleman et al. 2008); (2)~VLBI:~86~GHz (Lee et al. 2008); (3)~Power law fit of mm-data from NASA Extragalactic Database, evaluated at 1~mm; (4)~Submillimetre Common-User Bolometer Array (SCUBA):~2~mm (Di Matteo et al. 1999); (5)~Australia Telescope Compact Array (ATCA):~95~GHz (Sadler et al. 2008); (6)~NObeyama Bolometer Array (NOBA):~2~mm (Leeuw et al. 2004); (7)~Swedish-ESO 15m Submillimeter Telescope (SEST):~2~mm (Israel et al. 2008); (8)~SCUBA:~0.85~mm (Bendo et al. 2006); (9)~Very Long Baseline Array (VLBA):~8.4~GHz (Pellegrini et al. 2003); (10)~Plateau de Bure Interferometer (PdBI):~241.4~GHz (Sch\"odel et al. 2007); (11)~VLBI:~86~GHz (Middelberg et al. 2005). }
\end{deluxetable*}

In order to be able to resolve the photon ring with VLBI, the key question is whether the accretion flow of the target supermassive black hole is optically thin. In some cases, the spectra of these sources peak at wavelengths near $\lambda_{\rm max} \sim 1~{\rm mm}$, similar to the spectrum of Sgr~A*, suggesting that the emission comes from an ADAF (Di Matteo et al. 2000; Doi et al. 2005). Naively, we would expect an approximately linear scaling of the electron density of an ADAF with the ratio $M/\dot{M}$ of the black hole, where $\dot{M}$ is its mass accretion rate. The details of such a relation depend on a variety of factors, such as the temperature profile, the emissivity, and the radiative efficiency. However, most of the nearby supermassive black holes have very low radiative efficiencies (Ho 2009). Therefore, it is plausible that the accretion flows of these nearby supermassive black holes become optically thin at wavelengths that are comparable to 1~mm making them accessible to VLBI observations. Such observations are best carried out at wavelengths near the flux peak, where the accretion flow is becoming optically thin. At wavelengths $\lambda \ll \lambda_{\rm max}$, the emitted flux is likely to be too low to be detected with a VLBI array, while at wavelengths $\lambda \gg \lambda_{\rm max}$, the accretion flow is optically thick.

In Table~1, we summarize the angular diameters, distances, masses, radio luminosities $L_R$, flux densities $S_\nu$ near 1~mm, ratios of the bolometric luminosity to the Eddington luminosity $L_{\rm bol}/L_{\rm Edd}$, and peak wavelengths $\lambda_{\rm max}$ for supermassive black holes, whose photon rings have an angular diameter of at least $1~{\rm \mu arcsec}$.

In addition to Sgr~A*, the black holes in the centers of M87, M31, and M60 are good potential targets for VLBI observations, because of the large angular diameters of their respective photon rings and, in the case of M87 and M60, the measured peak in the synchrotron part of their spectra near 1~mm. M87 has a high measured flux density at 86~GHz (Lee et al. 2008) and should be readily observable at wavelengths close to 1~mm. In the case of M60, however, Di Matteo et al. (1999) report an upper limit on the flux density of $4~{\rm mJy}$.
No similar flux density measurement of M31 has been reported to date. We estimate the flux density at 1~mm of M31 from a simple power law fit of \mbox{mm-data} from the NASA Extragalactic Database. This flux density is relatively small, and M31 as well as M60 may be too faint to be observable with VLBI. Other sources, such as Centaurus~A, are luminous enough to be detectable at wavelengths near 1~mm. With increasing VLBI resolution, even their photon rings may become observable.

In the following, we assess the improvement on the mass measurements of the two supermassive black holes (Sgr~A* and M87) whose photon rings have the largest angular diameters. We assume fixed distances of $D_0=8.3~{\rm kpc}$ (Gillessen et al. 2009) and $D=17~{\rm Mpc}$ (G\"ultekin et al. 2009) for Sgr~A* and M87, respectively. For Sgr~A*, we estimate an error of the combined mass measurement from the existing stellar ephemerides and our simulated VLBI data of only $\sim 5\%$ (see Figure~\ref{contours}).

For M87, we estimate the smallest relative error that thermal noise-limited VLBI imaging observations of the ring can achieve from the signal to noise ratio for observations of Sgr~A* given by equation (\ref{SNR}), which we scale with the angular diameter of the black hole to obtain the expression
\begin{equation}
\frac{\delta M}{M} = \frac{\delta \theta_{\rm ring}}{\theta_{\rm ring}} \simeq \frac{1}{12} \lambda_{\rm mm}^{-2} \left( \frac{\theta_{\rm ring}}{21\lambda_{\rm mm}} - 1 \right)^{-1}.
\label{deltaM}
\end{equation}
Note, however, that M87 has a much larger mass and that the dynamical timescales of its accretion flow is much longer. Therefore, VLBI imaging observations of its photon ring will be much less affected by the variability of the accretion flow as in the case of Sgr~A*.

In Table~2, we compare the relative errors of the mass measurements of Sgr~A* and M87 (Gillessen et al. 2009; G\"ultekin et al. 2009) with our estimate of the error of VLBI observations of the respective photon rings at several different wavelengths. In the case of Sgr~A*, imaging its photon ring improves the error by a factor of about two. In the case of M87, imaging the photon ring at a wavelength of $0.5~{\rm mm}$ would lead to a result that is similar to the current mass measurement.

\begin{table}[h]
\begin{center}
\footnotesize
\begin{tabular}{lcccc}
\multicolumn{5}{c}{{\bf Table 2.} Relative Errors of Mass Measurements}\\
\hline \hline
Source & $\left. \frac{\delta M}{M} \right|_{\rm dyn}$ & $\left. \frac{\delta M}{M} \right|_{\rm VLBI}$ & $\left. \frac{\delta M}{M} \right|_{\rm VLBI}$ & $\left. \frac{\delta M}{M} \right|_{\rm VLBI}$\\
   & & (1.0 mm) & (0.8 mm) & (0.5 mm) \\
\hline
Sgr A* & 0.09 & 0.05 & 0.05 & 0.05 \\
M87 & 0.29 & 1.75 & 0.42 & 0.30 \\
\hline
\end{tabular}
\end{center}
\end{table}

As we pointed out in section~2 for the case of Sgr~A*, these errors require further refinement by in-depth imaging simulations. In addition, the morphology of \mbox{(sub-)mm} VLBI emission can be complicated by the presence of a jet (for M87, see Broderick \& Loeb 2009; Dexter et al. 2011).

As in the case of Sgr~A*, the combination of the results from both the dynamical and VLBI imaging observations of M87 would further reduce the error in the masses. The relative errors of mass measurements of both techniques likewise depend on the error in the measured distances to these sources. These errors, in turn, depend on uncertainties in the Hubble constant, peculiar motions of the gas in host galaxies, as well as assumptions on the proper motion of the Milky Way (see, e.g., Hodge 1981; Jacoby et al. 1992). The details of these effects on the relative errors of the mass are beyond the scope of our analysis.

\section{Discussion}

In this paper, we investigated the prospects of measuring the mass and distance of Sgr~A* as well as the masses of several other nearby supermassive black holes with a combination of dynamical observations and VLBI imaging of the respective photon rings of these sources. In order to resolve the photon ring of a black hole, its accretion flow must be optically thin. We argued that the wavelengths at which the accretion flows of these sources become optically thin should be roughly comparable to the location of the peak in the synchrotron emission of Sgr~A* and identified several supermassive black holes as optimal targets.

We explored the prospects of imaging the photon ring of Sgr~A* as well as of other nearby supermassive black holes with near-future VLBI arrays. We estimated the signal to noise ratio with which such arrays can image the photon ring in the best-case scenario if the VLBI observations are limited by thermal noise. Based on our estimate, we simulated confidence contours of a mass measurement of Sgr~A* using existing data of stellar ephemerides. We showed that the combination of both techniques can indeed reduce the correlation between mass and distance significantly resulting in relative errors of the mass and distance of only a few percent. We also identified several sources of uncertainty that have to be taken into account for an actual detection of the photon ring of Sgr~A*.

The uncertainties of measurements based on stellar orbits will be further reduced by the continued monitoring and by the expected improvement in astrometry possible with the second generation instrument {\it GRAVITY} for the {\it Very Large Telescope Interferometer} (Eisenhauer et al. 2011). Further improvements of the VLBI sensitivity will be achieved by the {\it Event Horizon Telescope}, a planned global array of \mbox{(sub-)mm} telescopes (Doeleman et al. 2009a,~b; Fish et al. 2009).

We estimated the improvement of the mass measurement of M87 using VLBI techniques. Such observations are promising at wavelengths near $0.5~{\rm mm}$ because of the large size of its photon ring. For M31 and M60, the supermassive black holes with the largest photon rings besides Sgr~A* and M87, the flux densities may be too low to be detectable with VLBI. As the resolution of VLBI arrays increases, additional sources will become observable.

Angular resolution of $\sim1~\mu$arcsec requires longer baselines and/or shorter observing wavelengths. However, the atmosphere precludes regular VLBI observations at wavelengths shorter than $\sim0.3~{\rm mm}$ at even the best sites.

A measurement of the photon-ring diameter is likely to yield useful results only if the observations extend well beyond the first null in the ring's visibility function. Even at a wavelength of $\lambda \simeq 0.4~{\rm mm}$, space-VLBI observations will be required to reach this point for all except the first four entries in Table~1.

VLBI between Earthbound antennas and a satellite has been achieved at 6~cm wavelength using the Japanese {\it HALCA} satellite (Hirabayashi et al. 1998). The recently launched {\it RadioAstron} (Kardashev 2009) will extend such observations to 1.2~cm wavelength and baselines as large as $4\times10^5~{\rm km}$, for a resolution of $<10$~$\mu$arcsec.

A future Explorer-class space mission designed to observe at 1~mm wavelength or shorter, where source opacity and scattering effects will be far less important, could provide the angular resolution needed to study a far larger sample of sources. Such a capability may be provided by the Russian-European {\it Millimetron} mission, which plans to deploy a 12~m antenna with VLBI capabilities to 0.4~mm and maximum baseline $>3\times10^5$~km (Wild et al. 2009).

We thank D. Zaritzky for useful comments. TJ and DP were supported by the NSF CAREER award NSF 0746549. This work was supported at MIT Haystack Observatory by NSF grants AST$-0908731$, AST$-0905844$, and AST$-0807843$.

\end{document}